\documentclass[a4paper,11pt]{article}
\pdfoutput=1

\usepackage{jinstpub}
\usepackage{booktabs}
\usepackage{subfigure}
\graphicspath{{figure/}}

\title{A LN$_{2}$-based cryogenic system prototype for future PandaX experiment}
\author[b]{Xinbei Jiang,}
\author[a,c,*]{Li Zhao,}
\author[a,b,*]{Shaobo Wang,}
\author[b]{Jia Fu,}
\author[b]{Tianjie Wu,}
\author[a,c,d]{Jianglai Liu,}

\affiliation[a]{INPAC, School of Physics and Astronomy, Shanghai Jiao Tong University,
             \\Shanghai Key Laboratory for Particle Physics and Cosmology,
             \\Key Laboratory for Particle Astrophysics and Cosmology, Ministry of Education (MoE), Shanghai 200240, China}
\affiliation[b]{SPEIT (SJTU Paris Elite Institute of Technology), Shanghai Jiao Tong University, Shanghai 200240, China}
\affiliation[c]{Shanghai Jiao Tong University Sichuan Research Institute, Chengdu 610213, China}
\affiliation[d]{Tsung-Dao Lee institute, Shanghai Jiao Tong University, Shanghai, 200240, China}

\emailAdd{zhaoli78@sjtu.edu.cn, shaobo.wang@sjtu.edu.cn}

\abstract{
This paper describes results on R$\&$D of an economical and efficient cryogenic system prototype for future liquid xenon detector.
The test module of the prototype has a "cold head" attached to a copper rod, which is specially designed to transport heat loads to a free-boiling liquid nitrogen bath.
The performance of the test module and commercial refrigerators is compared. The module with an optimized copper rod has demonstrated more than 1500~W cooling power at 178~K.
The temperature of the "cold head" can be kept stable with an error of 0.02~K, its fluctuation is within 0.1~K.
}

\begin{document}
\maketitle
\flushbottom
%\linenumbers

\section{Introduction}

The PandaX\cite{PandaX:2014ria, PandaX:2015gpz, PandaX:2016pdl,PandaX-II:2016vec,PandaX-II:2016wea,PandaX-II:2017hlx,Zhao:2020zy,PandaX-II:2020oim},  located in the China Jinping
Underground Laboratory (CJPL)~\cite{Kang:2010zza}, uses a dual-phase xenon time projection chamber to search for the dark matter particles in the form of weekly interacting massive particles (WIMPs)~\cite{Bertone:2004pz}.
The larger amount of the Liquid Xe~(LXe) target significantly increases the sensitivity of the experiment as the background is suppressed further.
Currently, the PandaX experiment has been upgraded to PandaX-4Twith $\sim$ 6 tons of xenon. The results of the commissioning run set the strongest upper limit~\cite{PandaX-4T:2021bab} on the dark matter-nucleon spin-independent interactions to date.
At the same time, XENONnT\cite{XENONnT:2021ldf} experiment with $\sim$ 8 tons of xenon is on commissioning, and LZ\cite{LZ:2019sgr} experiment with $\sim$ 10 tons of xenon is being constructed. The final phase of PandaX will be a detector with $\sim$ 50 tons of xenon,
it needs a more powerful cryogenic system.

In the past a few years, several different types of cooling systems have been used to support the operation of large LXe detectors.
PandaX-II\cite{Gong:2012thh} with $\sim$ 1.1 tons of xenon used a Pulse Tube Refrigerator (PTR) with a cooling power of $\sim$ 180~W at 178~K,
which penetrated into the LXe vessel with a copper finger; Now, PandaX-4T can run with three cooperating cold-heads, the total cooling power is $\sim$ 580~W at 178~K\cite{Zhao:2020vxh}.
XENONnT, using the same cryogenics of XENON1T\cite{Aprile:2012jh}, has 2 PTRs, the cooling power of each PTR is $\sim$ 250~W at 177~K. Unlike those above cooling systems with commercial refrigerators, the thermosyphon based cooling system with liquid nitrogen ($LN_{2}$) for LUX\cite{LUX:2012kmp} experiment,
which is also used for LZ~\cite{LZ:2017TDR}, has demonstrated more than 1000~W cooling power at 178~K. Another $LN_{2}-$based cooling system\cite{Giboni:2019fqo} without refrigerator has been discussed, however, there are no specific results.

In this paper, the results of R$\&$D on an economical and efficient cryogenic system for the future PandaX detector are reported.
The system is simple, cheap and free from mechanical vibrations.
The cooling power is from a bath of free-boiling $LN_{2}$. Heat conduction between "cold-head" and $LN_{2}$ is provided with a special designed Oxygen-Free High-Conductivity (OFHC) copper rod.
The system with different size rods demonstrates different cooling power, which is also useful for other cryogenic experiments.

\section{LN$_{2}$ based cryogenic system prototype}
\subsection{Principle of operation}
\begin{figure}[h]
    \centering
    \includegraphics[width = 0.8\textwidth]{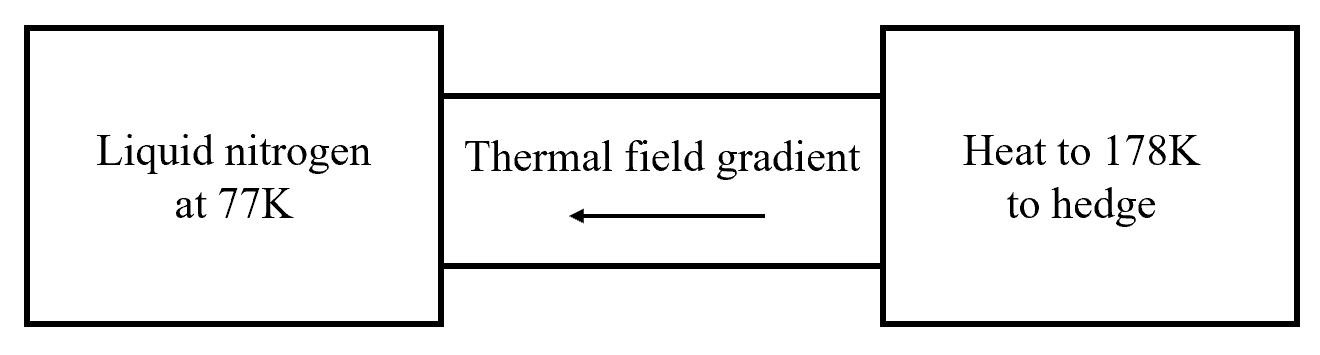}
    \caption{The simplified model with a copper conductor between $LN_{2}$ and $LXe$.}
    \label{fig:model}
\end{figure}

A specially designed cooling system mainly includes $LN_{2}$ (77~K) , $LXe$ (178~K) and a thermal conductor between them.
According to its high degree of symmetry, the cooling system is simplified to an one-dimensional model based on heat conduction, presented in Figure \ref{fig:model}.
The copper rod with high thermal conductivity, as a cylindrical bar, is connected to the cold source $LN_{2}$ at one end, and connected to the heat load $LXe$  at the other end.
Driven by the heat conduction theory, a considerable temperature difference between the two ends leads to a certain heat transfer power through the copper rod.

The Fourier's law applied to the model at steady state follows:
\begin{equation}
    P_{cond}=\lambda A \frac{T_{LN_{2}}-T_{LXe}}{L}
     \label{eq:diffusion}
\end{equation}
where $P_{cond}$ is the cooling power through the copper rod, $A$ is cross section of the copper rod, $L$ is length of the copper rod, $\lambda$ is thermal conductivity of the copper, $T_{LN_{2}}$ is temperature of $LN_{2}$, $T_{LXe}$ is temperature of $LXe$.

The thermal conductivity of oxygen-free copper is $\sim 400$~W/(m$\cdot$K) \cite{cond:Cu}. Total heat leakage of the system is also estimated, including radiation and convection.
Two kinds of the model are calculated here: 1) 200~W@178~K, almost the same to that of a normal Pulse Tube Refrigerator (PTR) PC-150\cite{PTR:PC-150} (JEC, Japan); 2) 1500~W@178~K,
like a most powerful single stage Gifford McMahon (GM) refrigerator AL600\cite{GM:AL600} (Cryomech, U.S.A).

Based on the Fourier's law, preliminary ideal sizes of copper rod are presented in Table \ref{tab:para}.
\begin{table}[ht]
	\centering
	\caption {Preliminary ideal sizes of the copper rod}
	\label{tab:para}
	\begin{tabular}[c]{cccc}
		\toprule
		{Effective cooling power(W)} & {Heat Leakage(W)} & {Length(mm)} & {Diameter(mm)}  \\
		\midrule
		$200$ & $30$ & $124$ & $30$   \\
		$1500$ & $100$ & $127$ & $80$  \\
		\bottomrule
	\end{tabular}
\end{table}

\subsection{Test module}\label{des}
\begin{figure}
    \centering
    \includegraphics[width = 0.8\textwidth]{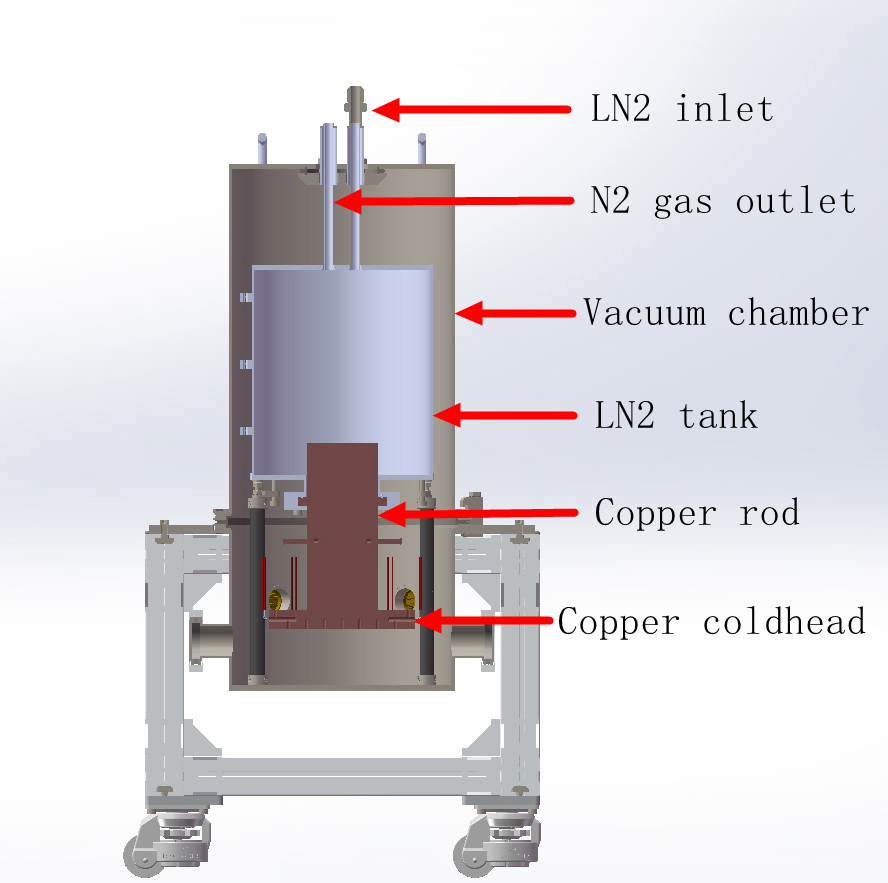}
    \caption{The section view of the test module}
    \label{fig:proto}
\end{figure}

\begin{figure}
  \centering
  \subfigure[Copper rod for 200~W cooling power]{\includegraphics[width = 0.391\textwidth]{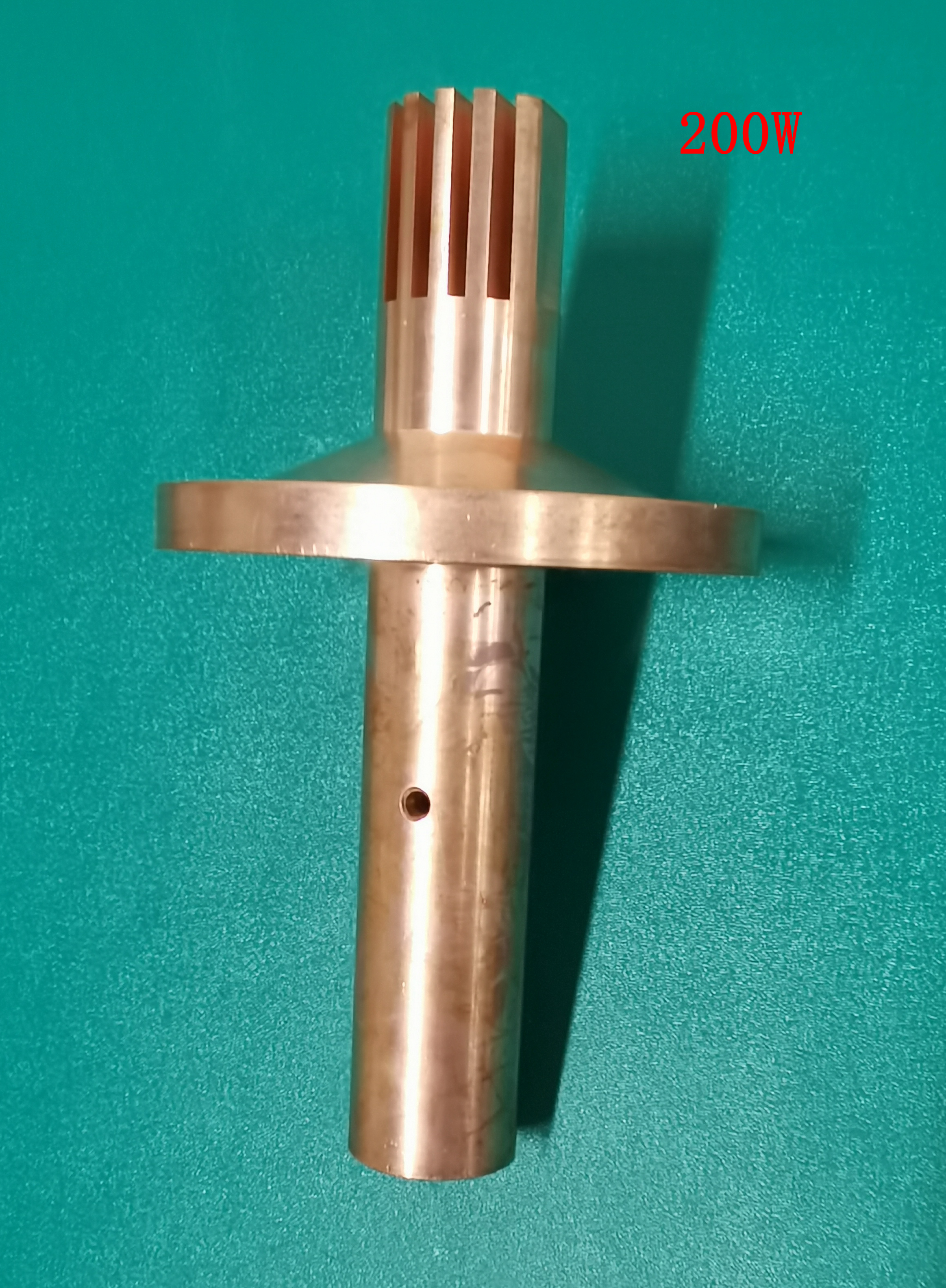}}
  \subfigure[Copper rod for 1500~W cooling power]{\includegraphics[width = 0.4\textwidth]{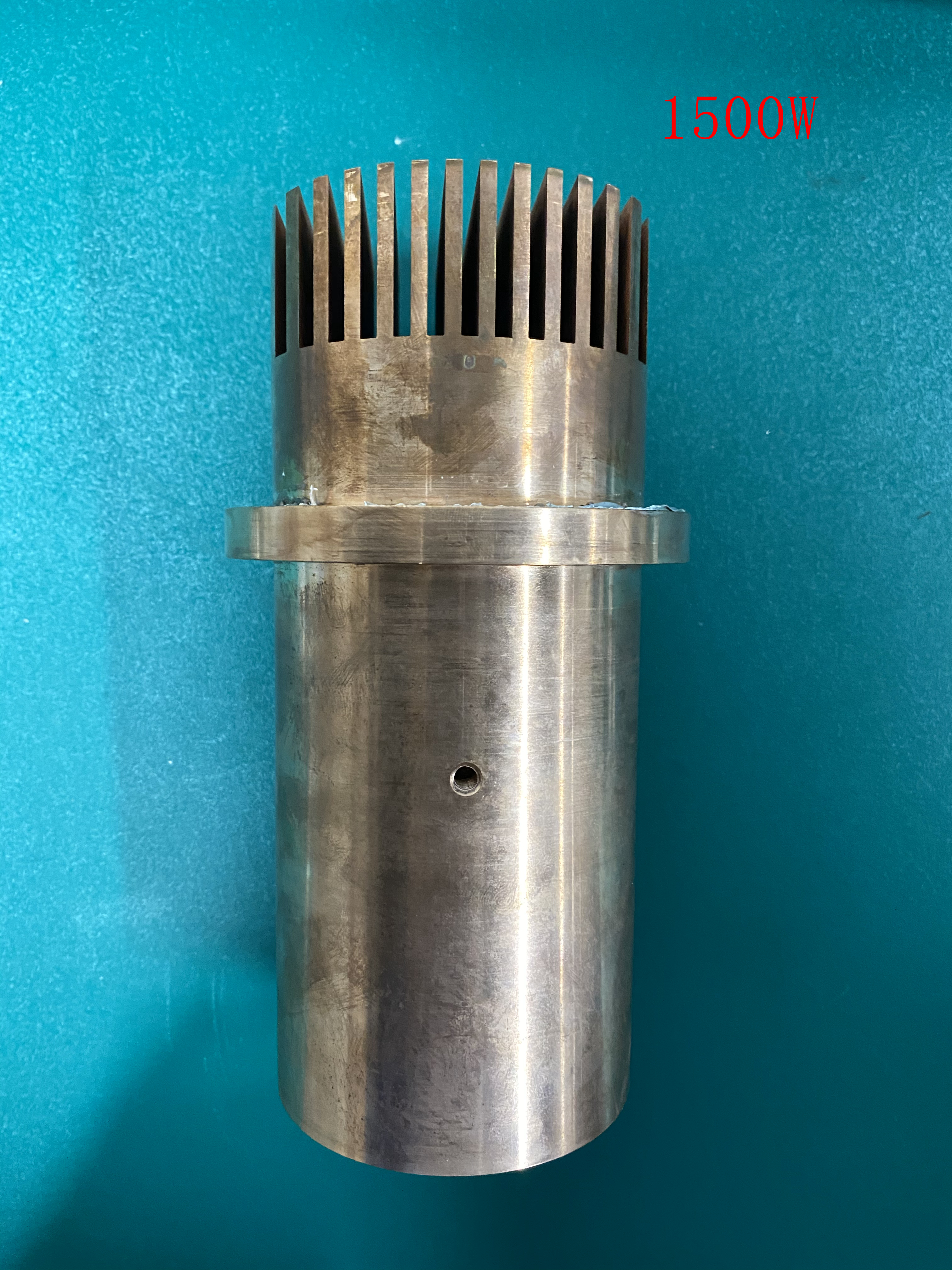}}
  \caption{Pictures of the copper rods as the conductor}
  \label{fig:copper-rod}
\end{figure}

A particular test module (Figure \ref{fig:proto}) is designed, which mainly consists of a vacuum chamber, a $LN_{2}$ tank,
a copper rod and a copper "coldhead". A 165~mm diameter copper disc with the thickness of 22~mm is chosen to be a heat load, as a common "coldhead" for two kinds of modules. Using cryogenic thermal grease, Pt100 sensors and cartridge heaters (10$\times$15~ W) are inserted into suitable holes of the copper disc. The "coldhead" is attached tightly to the bottom section of the copper rod (Figure \ref{fig:copper-rod}) with 0.3~mm thickness indium sheet and screws (4$\times$M6).
The top fin of the copper rod is immersed in $LN_{2}$, with larger heat exchange area. In order to prevent $LN_{2}$ from leaking into the vacuum chamber,
the 10~mm thickness raised lateral ring of the copper rod is fixed to the bottom flange of the $LN_{2}$ tank with 2~mm diameter indium wire.
The $\sim$7~$l$ tank with a $LN_{2}$ inlet (diameter: 9~mm) and $N_{2}$ gas outlet (diameter: 9~mm) are located in upper space of the vacuum chamber, its volume is large enough for timely $LN_{2}$ replenishment.
It is supported by three glass fiber reinforced plastics feet, which stand on the bottom plate of the vacuum chamber.

Finally, all the low temperature parts are enclosed in the 250~mm diameter vacuum chamber (total height: $\sim$ 600~mm) with aluminized Mylar foil and high reflective film to minimize heat leaks from the outside walls.
The vacuum chamber would be connected to a set of pumps during the experiment.

\subsection{Experimental setup}

\begin{figure}
    \centering
    \includegraphics[width = 0.8\textwidth]{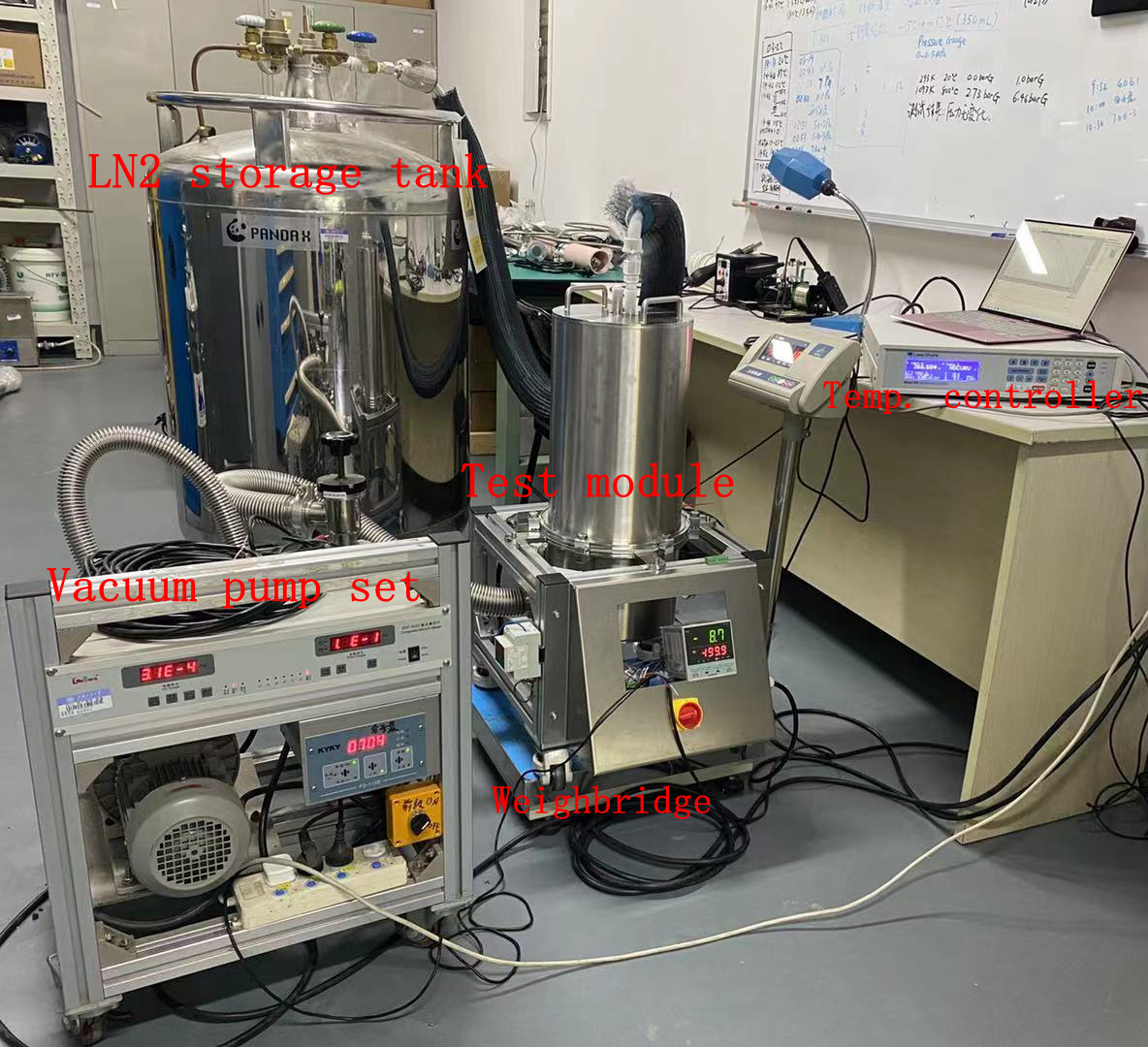}
    \caption{Equipment assembly}
    \label{fig:assembly}
\end{figure}

As shown in Figure \ref{fig:assembly}, the experimental setup mainly contains five parts: a vacuum pump set, a 200~$l$ $LN_{2}$ storage tank, a temperature controller, a weigh-bridge and the test module.
The chamber of the test module is pumped by the vacuum pump set for getting good insulation vacuum. The test module is put on the weigh-bridge, its total weight can be measured, then,
the left $LN_{2}$ in inside tank after partial evaporation can be monitored, which is useful for timely refilling $LN_{2}$ and calculating $LN_{2}$ consumption.
The temperature of the "coldhead" in the test module is regulated by the temperature controller (Lakeshore 350 \cite{controller:350}) and the heaters in the "coldhead".

\subsection{Test procedure}
At the beginning of the experiment, the vacuum chamber is pumped overnight at room temperature. The effective cooling power of the test module will be measured as the vacuum chamber in a good vacuum($<1\times10^{-3}$ Pa).
The Lakeshore 350 Proportional, Integral and Derivative (PID) controller reads the Pt100 sensors in the "coldhead" and controls the electric power to the heaters. Considering the limit of the maximum power (75~W) of heating of the controller,
the secondary control loop of 350, providing a $0\sim10$ V DC signal, controls a single phase intergraded phase-shift trigger module power supply of the AC heaters, the temperature of the "coldhead" can be stable at the set value.

Before formal measurement, $LN_{2}$ is filled into the inside tank and stopped as the liquid amount is $\sim2/3$ of full. At the same time, the temperature of the "cold-head" will go down and be stable at the lowest temperature, the vacuum is getting better ($\sim 5\times10^{-5}$ Pa).
Then, the heat leakage of the test module at this case can be calculated by recording the $LN_{2}$ consumption over time.

Finally, the effective cooling power of the test module at different temperature set-point would be measured. In order to check its stability and reliability, every experiment would be carried out in cycles (cooling down and warming up) at least three times.

\section{Results and discussion}

After cooling inside parts down enough, the heat leakage ($H_{leak}$) is estimated using the consumption of the LN2 as no heating.
The rate of $LN_{2}$ consumption of 200~W@178~K and 1500~W@178~K copper rods is $120\pm40$~g/h and $300\pm100$~g/h respectively.
\begin{equation}
    H_{leak}=h\times \upsilon /M
\end{equation}
where $h$ is the latent heat of vaporization of $LN_{2}$ (5.56~KJ/mol), $\upsilon$ is the rate of $LN_{2}$ consumption and M is Molar quantity of $N_{2}$ (28~g/mol).
Therefore, the heat leakage of 200~W@178~K and 1500~W@178~K test module is $7\pm3$ W and $17\pm6$ W respectively.

\subsection{Performance of the 200~W@178~K test module}
\begin{figure}[h]
  \centering
  \includegraphics[width = 0.8\textwidth]{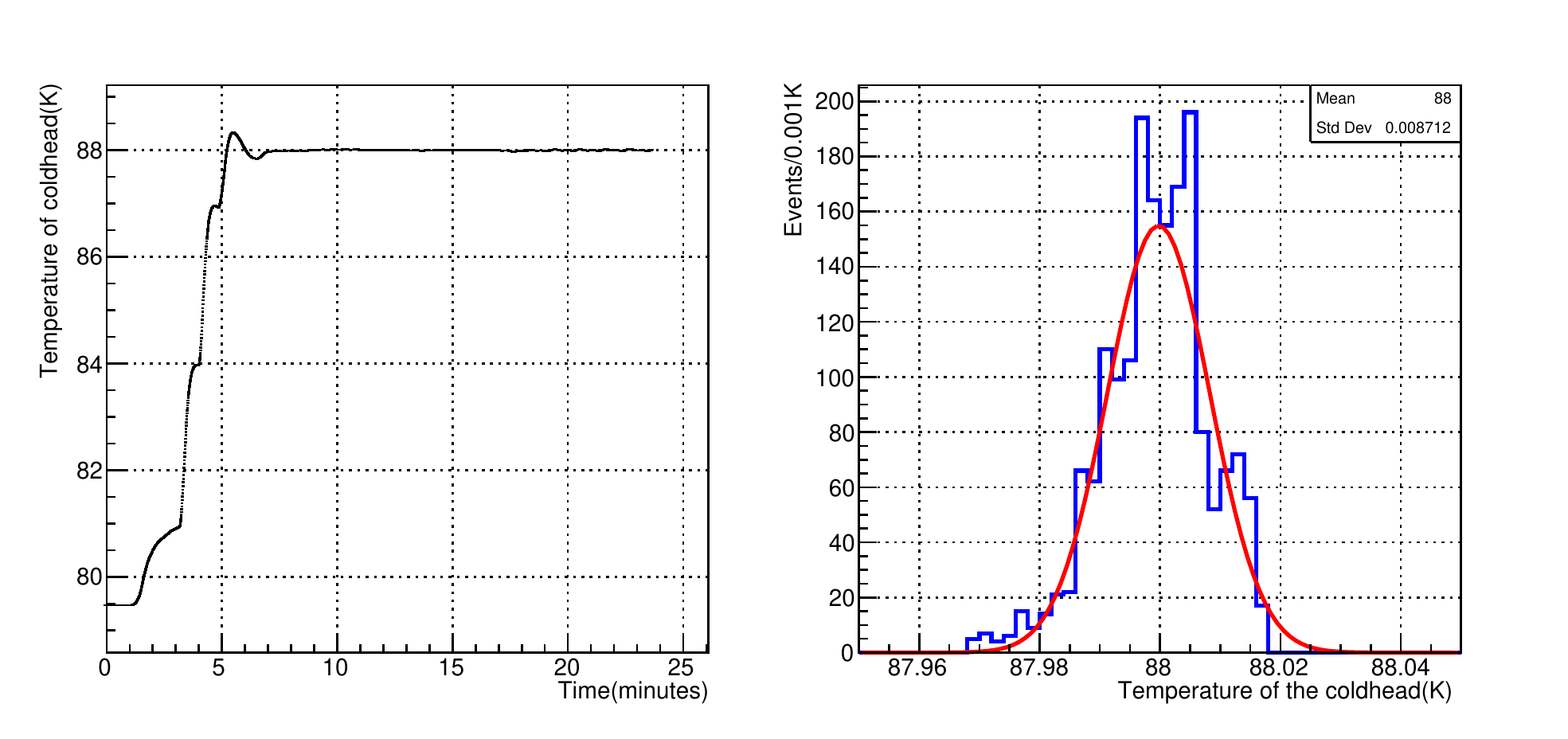}
  \includegraphics[width = 0.8\textwidth]{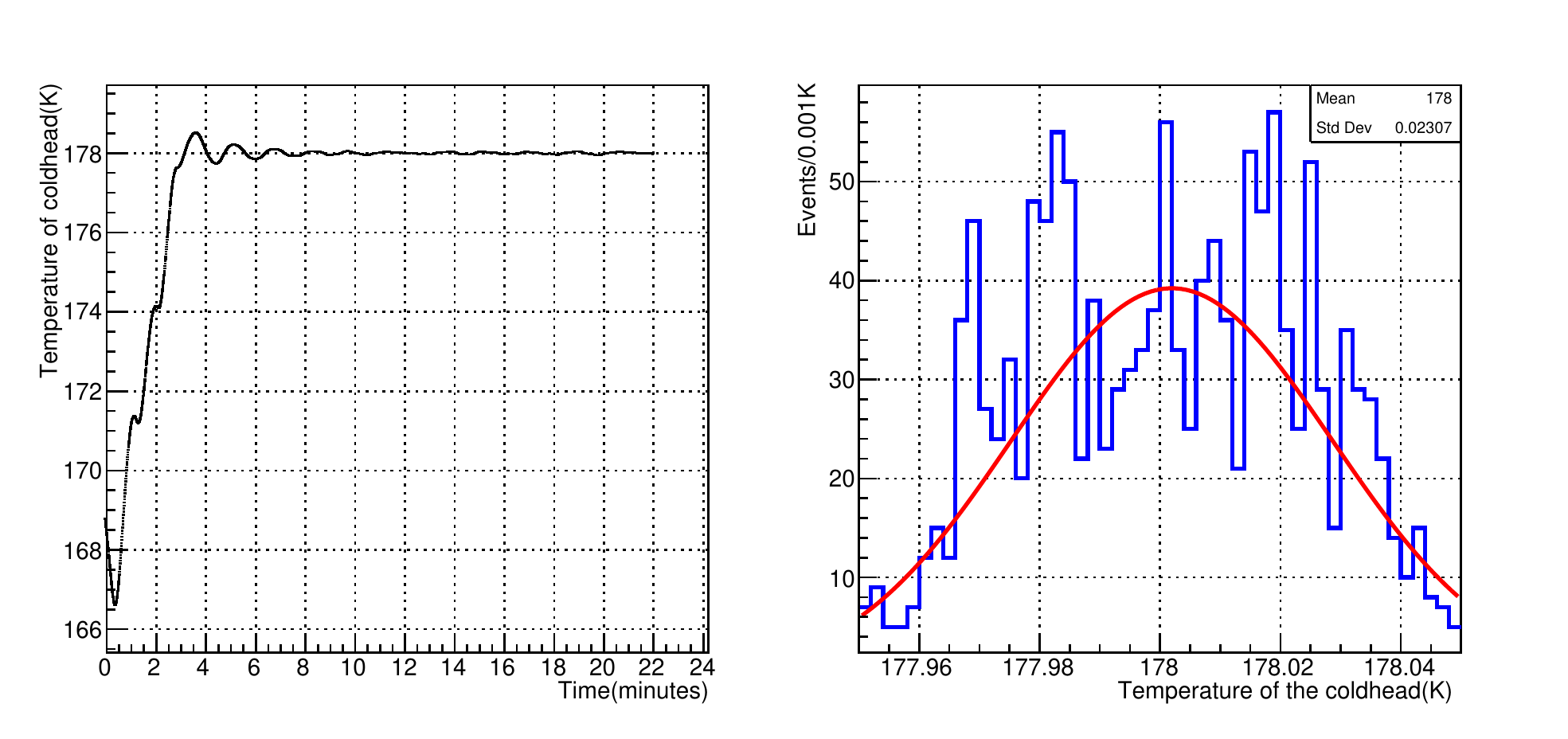}
  \caption{Temperature evolution and fluctuation at 88~K (top) and 178~K (bottom) for the 200~W@178~K copper rod, the distribution is fitted with a Gaussian function}
  \label{fig:temp-200W}
\end{figure}
The 200~W@178~K test module was tested, the temperature of the "coldhead" was $78.31\pm0.02$~K after 1.5 hours cooling with $LN_{2}$,
it was about 1~K higher than the temperature of $LN_{2}$. Then, some tests of the effective cooling power were carried out by heating the "coldhead" gradually.
Temperature evolution and fluctuation at two typical set-points (88~K, 178~K) is shown in Figure \ref{fig:temp-200W}. The temperature of the "cold-head" can be stable in 10 minutes after it reaches the set-point, its standard deviation at stable case is $\sim$ 0.02 K.

\begin{figure}
    \centering
    \includegraphics[width = 0.7\textwidth]{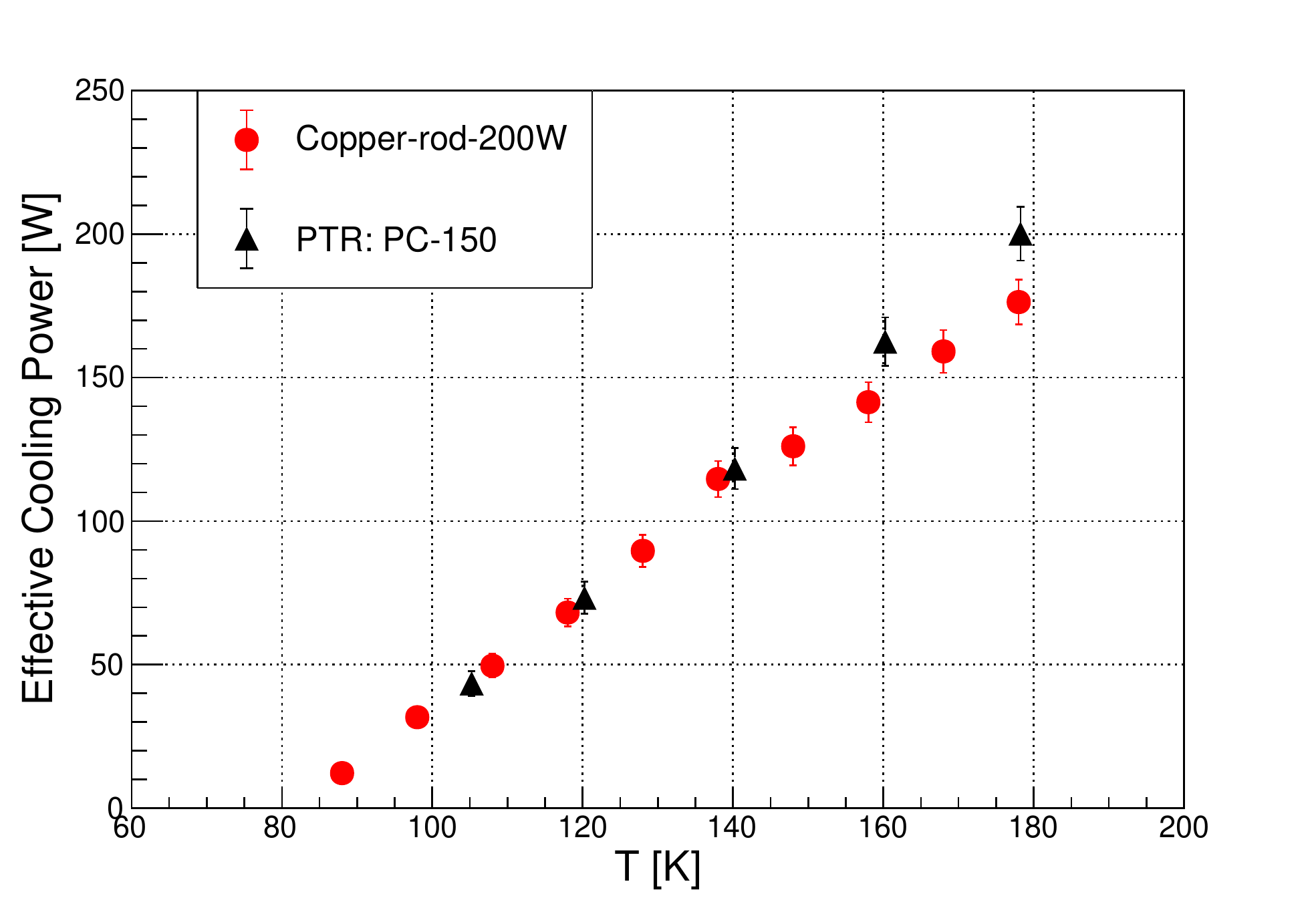}
    \caption{The effective cooling power of the 200~W@178~K copper rod and PTR PC-150\cite{Zhao:2020vxh}}
    \label{fig:cooler-200W}
\end{figure}

The test results are shown in Figure~\ref{fig:cooler-200W}. The results of PTR PC-150 for PandaX-4T experiment are plotted too. Their cooling power below 140~K is almost the same.
However, the cooling power of the 200~W@178~K copper rod becomes smaller than that of PC-150 as the temperature is greater than 140~K, its slope decreases.
The equation~\ref{eq:diffusion} of simplified module is for $LN_{2}$ steady state, there is a linear relationship between the cooling power and temperature set-point ($T_{LXe}$) once the setup is fixed. Nevertheless,
$LN_{2}$ insides will change to superheated liquid under $>1.0$ bar pressure as a 9~mm diameter limited outlet of exhausting $N_{2}$ gas is used and the heating power goes up,
which means that $T_{LN_{2}}$ will be greater than 77~K. Besides, actual thermal conductivity between the "cold-head" and $LN_{2}$ using the copper rod and the indium sheet is $379\pm8$ W/(m$\cdot$K) as considering the heat leakage and $4\times M6$ screw holes, a little lower than that of pure copper.
Therefore, the cooling power of the test module is less than expected one at 178~K.

\subsection{Performance of the 1500~W@178~K test module}
\begin{figure}
    \centering
    \includegraphics[width = 0.7\textwidth]{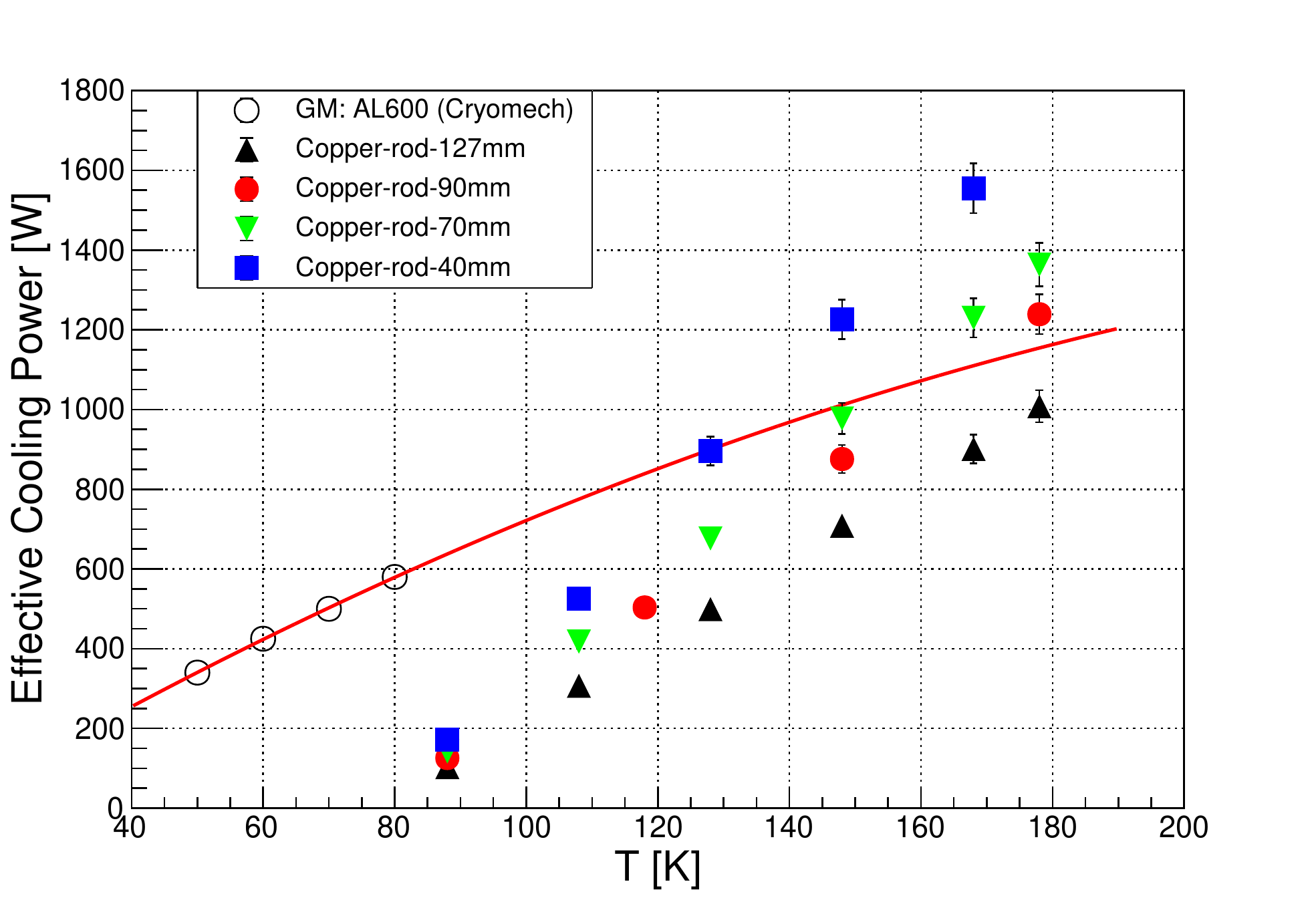}
    \caption{The effective cooling power of the 1500W@178K copper rods (127 mm, 90 mm, 70 mm, 40 mm) and GM AL600\cite{GM:AL600}, the data points of AL600 from Cryomech company are fitted with a quadratic polynomial function.}
    \label{fig:cooler-1500W}
\end{figure}

The 200~W@178~K copper rod was replaced with 1500~W@178~K ones, other parts were kept. Then, some tests of the effective cooling power with different copper rods were carried out by heating the "cold head" gradually.
The test results are shown in Figure \ref{fig:cooler-1500W}. The data points of GM Al600 from the company are plotted too, a quadratic polynomial function is used to fit the data.

Several copper rods (127~mm, 90~mm, 70~mm, 40~mm) are tested, their cooling power increases slowly with shorter length ($L$) and is nonlinear to the parameter $\frac{1}{L}$, which is unexpected. There are two disadvantageous factors for this test module as the following: 1) The superheated effect ($T_{LN_{2}}>$77~K) for this test module becomes more obvious as the heating is higher and higher; 2) 4*M6 screws on small circle (20 mm diameter) are a little less for fixing the big rod (80 mm diameter) and "the cold head", which results in uneven stress on contacted surfaces and lower effective thermal conductivity ($\lambda$). However, the test module with short copper rods (90~mm, 70~mm, 40~mm) is more powerful at 178~K than that of AL600 commercial refrigerator, which shows good linearity. The cooling power of the copper rod (length: 40~mm, diameter: 80~mm) can be greater than 1500~W at 178~K.

\begin{figure}[h]
  \centering
  \includegraphics[width = 0.8\textwidth]{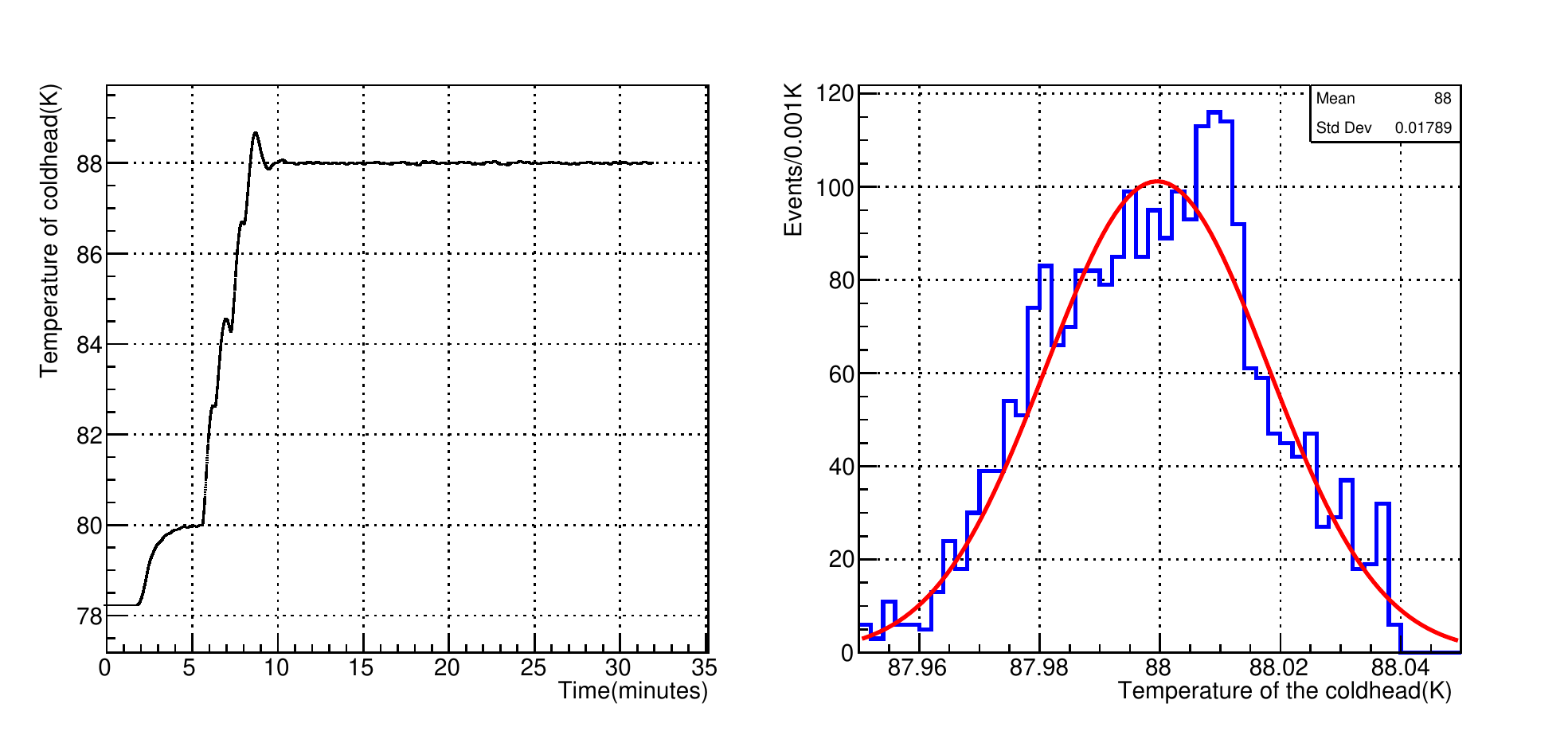}
  \includegraphics[width = 0.8\textwidth]{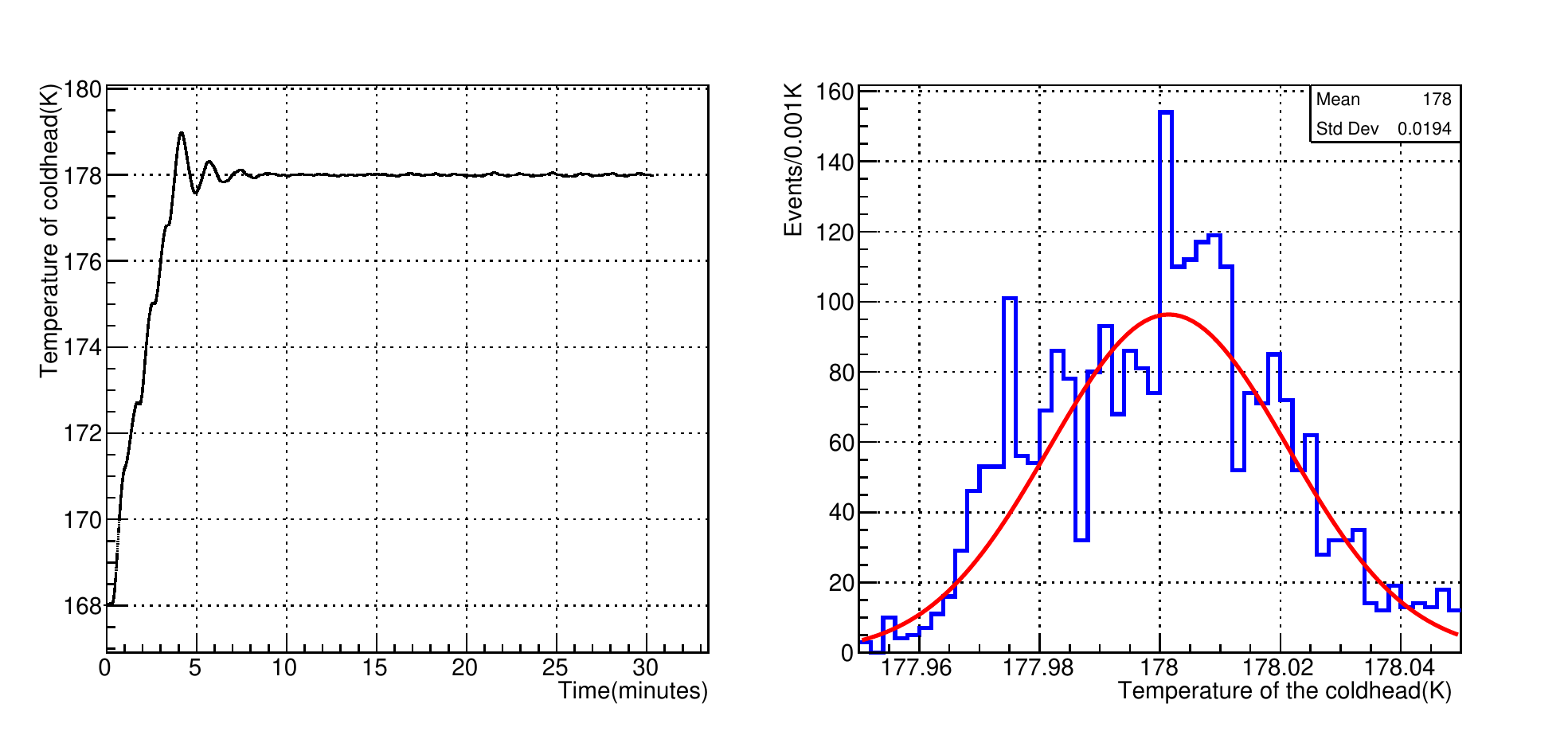}
  \caption{Temperature evolution and fluctuation at 88~K (top) and 178~K (bottom) for the 1500W@178K copper rod (diameter: 80~mm, length: 127~mm), the distribution is fitted with a Gaussian function}
  \label{fig:temp-1500W}
\end{figure}

Temperature evolution and fluctuation of 1500~W@178~K copper rod is shown in Figure \ref{fig:temp-1500W}. It is similar to that of 200~W@178~K one. The standard deviation of temperature is still $\sim$ 0.02~K for powerful test modules even for larger voltage output fluctuation of the heating power supply.

Therefore, this $LN_{2}$ based cooling system shows high cooling power and stability at 178~K, it will be tested using Xenon gas in the future.

\section{Conclusion}
In this paper, the cooling system prototype based on $LN_{2}$ and copper thermal conductor has been designed and experimentally investigated.
The experiment shows that the prototype can provide a cooling power of $>$ 1500~W at 178~K using an optimized copper rod (diameter: 80~mm, length: 40~mm). The stable operation temperature has an error of $\sim$ 0.02~K,
its fluctuation is within 0.1~K. Besides, the system exhibits high reliability and repeatability. In a word, it will be a potential replacement of conventional refrigerator for large LXe detector in future.

\acknowledgments
This project is supported in part by a grant from the Ministry of Science and Technology of China (No. 2016YFA0400301) and a grant from National Science Foundation of China (No. 12090061). We thank supports from Double First Class Plan of the Shanghai Jiao Tong University. We also thank the sponsorship from the Hongwen Foundation in Hong Kong, and Tencent Foundation in China.

%\section*{References}
%\bibliographystyle{unsrt}
%\bibliography{mybibfile}

\end{document}